\documentclass{elsart}
\usepackage{graphicx}
\def\Vec#1{\mbox{\boldmath $#1$}}
\input epsf.sty

\usepackage{amssymb}

\begin{document}
\begin{frontmatter}
\title
{ An Analysis of Chaos via \\Contact Transformation }
\author
{Sadataka Furui} 
\address
{School of Science and Engineering, Teikyo University, \\320-8551, Utsunomiya, Japan}
\ead{furui@umb.teikyo-u.ac.jp}

\begin{abstract}
Transition from chaotic to quasi-periodic phase in modified Lorenz model is analyzed by performing the contact transformation such that the trajectory in ${\Vec R}^3$ is projected on ${\Vec R}^2$.  The relative torsion number and the
characteristics of the template are measured using the eigenvector of the Jacobian instead of vectors on moving frame along the closed trajectory. 

Application to the circulation of a fluid in a convection loop and oscillation of the electric field in single-mode laser system are performed. The time series of the eigenvalues of the Jacobian  and the scatter plot of the trajectory in the transformed coordinate plane $X-Z$ in the former and $|X|-|Z|$ in the latter,  allow to visualize characteristic pattern change at the transition from quasi-periodic to chaotic. In the case of single mode laser, we observe the correlation between the critical movement of the eigenvalues of the Jacobian in the complex plane and intermittency.
\end{abstract}

\begin{keyword}
Chaos \sep Intermittency \sep Contact transformation
\PACS 5.45 \sep 47.52 \sep 47.27 
\end{keyword}
\end{frontmatter}

\section{Introduction}
In dissipative dynamical systems exhibiting transition from periodic or quasi-periodic behaviour to chaotic one, topological method is a useful tool. Gilmore\cite{Gil} has recently reviewed how the integer invariants can be obtained from chaotic time series. In the time series, sudden bursting is called intermittency. Pomeau and Manneville\cite{PM,Pom} proposed three types of routes to chaos which accompany intermittency. The type I is due to disappearance of a stable periodic orbit via pitchfork bifurcation, the type II is via period doubling bifurcation and the type III is via Poincar\'e-Andronov-Hopf bifurcation. The definitions of the type of bifurcation can be seen in e.g.\cite{MoAr,Wig}. 

In the Lorenz model, which simulates the flow of fluids between hot lower plate and cold upper plate, chaotic behaviour arises at a certain parameter region through the pitchfork bifurcation.
Birman and Williams\cite{BW} showed that the strange attractor of the Lorenz model in ${\Vec R}^3$ can be constructed on a 2-dimensional bifurcating manifold which is called knot-holder or a template, and that the topological analysis of the orbits can be reduced to that of braid groups. 

By applying external temperature oscillation $\rho_1\cos\omega t$ on the upper plate, one can stabilize the chaotic behaviour in the Lorenz model. In this case the dynamics is extended from ${\Vec R}^3$ to ${\Vec R}^3\times S^1$ where $S^1$  is the 1-dimensional torus, and one can guide the trajectory of the orbit from chaotic to periodic or quasi-periodic one. 

Holmes and Williams\cite{HW} extended the Lorenz mapping to mappings that contain Smale's horseshoe mapping which is characterized by stretching and folding, and expressed dynamics on the horseshoe template as a sequence of symbols L and R corresponding to the relative position of the two orbits on the template.  They showed that the period-doubling bifurcations can be specified by the number of half twist $\xi$ and the index of folding $p$ and called the associated 2-dimensional manifold as  $(\xi, p)$-template.

Aizawa and Uezu\cite{AiUe,UeAi,Ue} performed the topological analysis of the perturbed Lorenz model and specified the bifurcating orbits by the linking number of two orbits $L(C_1,C_2)$, torsion number $n_i$ and relative torsion number $r_i$ which is equivalent to $p/2$ of the Holmes-Williams, and derived a recursion formula for those of the $2^k$th bifurcation orbit $r_i^{(k)}$ and $n_i^{(k)}$.

Arimitsu and Motoike\cite{MoArKo,ArMo,MoAr} extended the analysis of Holmes and Williams and showed that from the power spectrum of the bifurcating orbits, topological properties of the orbit can be extracted. They defined the local crossing number $C_{\langle n\rangle}$, global crossing number $c_{\langle n\rangle}$ and the linking number $l_{\langle n,n-1\rangle}$ of the $2^n$th and the $2^{n-1}$the bifurcation orbit and obtained the recursion formula. 

The chaos control of the perturbed Lorenz system was studied from the technological point of view\cite{Dor,Cre}. By a simple $\rho_1\cos \omega t$ type parametric driving of a convection loop model, it was shown that the chaotic system after driven to periodicity can be led back to chaotic behaviour through intermittency. General methods of chaos control are proposed by several authors\cite{OGY,Pyr}. 

There are analogies between the Lorenz model and the single mode laser system\cite{Hak,Hakb}. The dynamics of the electric field, polarization and the inversion density can be assigned as the coordinate of the three dimensional manifold, but the electric field and polarization are accompanied by the modulation, and they are expressed as complex numbers. The transition to chaos by applying external constant electric field to the single mode laser system was simulated\cite{Lug}.

In this paper, we analyze the convection loop model and the single mode laser system from a slightly different point of view. We adopt the framework of differential geometry and adopt the contact transformation to simplify the analysis. We consider the topological properties of these models following\cite{UeAi,ArMo} and consider an extension of our method to other systems.

\section{Contact transformation}
When an orbit in a d-dimensional dynamical system is specified by
$t,x_1,x_2,\cdots,x_d,$ 
$p_1,p_2,\cdots,p_d$, and if the coordinate transformation
\begin{equation}
T=T(t,x_1,x_2,\cdots,x_d,p_1,p_2,\cdots,p_d)
\end{equation}
\begin{equation}
X_j=X_j(t,x_1,x_2,\cdots,x_d,p_1,p_2,\cdots,p_d)
\end{equation}
\begin{equation}
P_j=P_j(t,x_1,x_2,\cdots,x_d,p_1,p_2,\cdots,p_d)
\end{equation}
does not change the total differential equation
\begin{equation}
dt-p_1dx_1-p_2dx_2-\cdots-p_d dx_d=0\label{diff}
\end{equation}
i.e. when the equation
\begin{eqnarray}
&&dT-P_1dX_1-P_2dX_2-\cdots-P_d dX_d\nonumber\\
&=&\rho(t,x_1,x_2,\cdots,x_d,p_1,p_2,\cdots,p_d)(dt-p_1dx_1-p_2dx_2-\cdots-p_d dx_d)\nonumber\\
&=&0
\end{eqnarray}
is identically satisfied, the transformation is called contact transformation.

When the dynamical equation is given as
\begin{eqnarray}
\frac{d x}{d t}&=&h_1(x,y)\nonumber\\
\frac{d y}{d t}&=&h_2(x,y,z)\nonumber\\
\frac{d z}{d t}&=&h_3(x,y,z)
\end{eqnarray}  
we perform the transformation $X=x, Y=h_1(x,y)$. By using $\displaystyle \frac{d x}{d t}=h_1(x,y)$ and $\displaystyle \frac{d y}{d t}=h_2(x,y,z)$, the derivative $\displaystyle \frac{d Y}{dt}$ can be expressed as $g_2(X,Y,z)$, and we define it as $Z$. Further using $\displaystyle \frac{d z}{d t}=h_3(x,y,z)$, the derivative
$\displaystyle \frac{d Z}{dt}$ can be expressed as $f(X,Y,Z)$. Thus the system can be expressed as
\begin{eqnarray}
\frac{d X}{d t}&=&Y\nonumber\\
\frac{d Y}{d t}&=&Z\nonumber\\
\frac{d Z}{d t}&=&f(X,Y,Z)
\end{eqnarray} 
To show that it is a particular case of contact transformation, it is sufficient to consider the identity
\begin{eqnarray}
&&Z f(X,Y,Z)dX+Y f(X,Y,Z)dY+Y Z dZ\nonumber\\
&=&\frac{Y Z f(X,Y,Z)}{h_1(x,y)h_2(x,y,z)h_3(x,y,z)}\nonumber\\
&&\times (h_2(x,y,z)h_3(x,y,z) dx+h_1(x,y)h_3(x,y,z) dy+h_1(x,y)h_2(x,y,z) dz) \nonumber\\
&=&\rho(x,y,z)(h_2(x,y,z)h_3(x,y,z) dx+h_1(x,y)h_3(x,y,z) dy\nonumber\\
&&+h_1(x,y)h_2(x,y,z) dz)
\end{eqnarray}

When the system is discretized we obtain
\begin{eqnarray}
X_{n+1}&=&X_n+Y_n dt\nonumber\\
Y_{n+1}&=&Y_n+Z_n dt\nonumber\\
Z_{n+1}&=&Z_n+ f(X_n,Y_n,Z_n)dt
\end{eqnarray}

The Jacobian of this mapping is
\begin{eqnarray}
&&\left(
\begin{array}{c c c}
 \frac{\partial (X_{n+1}-X_n)}{\partial X_n} & \frac{\partial (X_{n+1}-X_n)}{\partial Y_n} & \frac{\partial (X_{n+1}-X_n)}{\partial Z_n} \\
 \frac{\partial (Y_{n+1}-Y_n)}{\partial X_n} & \frac{\partial (Y_{n+1}-Y_n)}{\partial Y_n} & \frac{\partial (Y_{n+1}-Y_n)}{\partial Z_n} \\
 \frac{\partial (Z_{n+1}-Z_n)}{\partial X_n}&\frac{\partial (Z_{n+1}-Z_n)}{\partial Y_n}& \frac{\partial (Z_{n+1}-Z_n)}{\partial Z_n}\end{array}
\right)/dt\nonumber\\
&&=
\left(
\begin{array}{c c c}
 0 & 1 & 0\\
 0 & 0 & 1\\
 \frac{\partial f(X_{n+1},Y_{n+1}, Z_{n+1})}{\partial X_n}&\frac{\partial f(X_{n+1},Y_{n+1},Z_{n+1})}{\partial Y_n}&\frac{\partial f(X_{n+1},Y_{n+1},Z_{n+1})}{\partial Z_n}\end{array}
\right).\label{jacobi}
\end{eqnarray}
and its eigenvalues are given by the solution of
\begin{eqnarray}
&&\lambda^3- \frac{\partial f(X_{n+1},Y_{n+1}, Z_{n+1})}{\partial X_n}\lambda^2
-\frac{\partial f(X_{n+1},Y_{n+1},Z_{n+1})}{\partial Y_n}\lambda
\nonumber\\
&&-\frac{\partial f(X_{n+1},Y_{n+1},Z_{n+1})}{\partial Z_n}=0,
\end{eqnarray}
which means that there is a zero eigenvalue if $\displaystyle\frac{\partial f(X_{n+1},Y_{n+1},Z_{n+1})}{\partial Z_n}=0$ is satisfied. 

\section{Lorenz equation}

In the case of Lorenz equation
\begin{eqnarray}
\frac{d x}{d t}&=&\sigma(y-x)\nonumber\\
\frac{d y}{d t}&=&-y+r x-xz\nonumber\\
\frac{d z}{d t}&=&-bz+xy\label{mlor3}
\end{eqnarray}

The derivative $\displaystyle\frac{dZ}{dt}=f(X,Y,Z)$  becomes
\begin{eqnarray}
f(X,Y,Z)&=&(YZ+\sigma Y^2+Y^2-\sigma X Z-X Z-X^3Y-\sigma X^4\nonumber\\
&&-bX Z-\sigma b X Y+\sigma b r X^2-b X Y-\sigma b X^2)/X, 
\end{eqnarray}

\subsection{Scatter plot in the $X-Z$ plane}
\begin{figure}[e]
\begin{minipage}[b]{0.47\linewidth} 
\begin{center}
\epsfysize=100pt\epsfbox{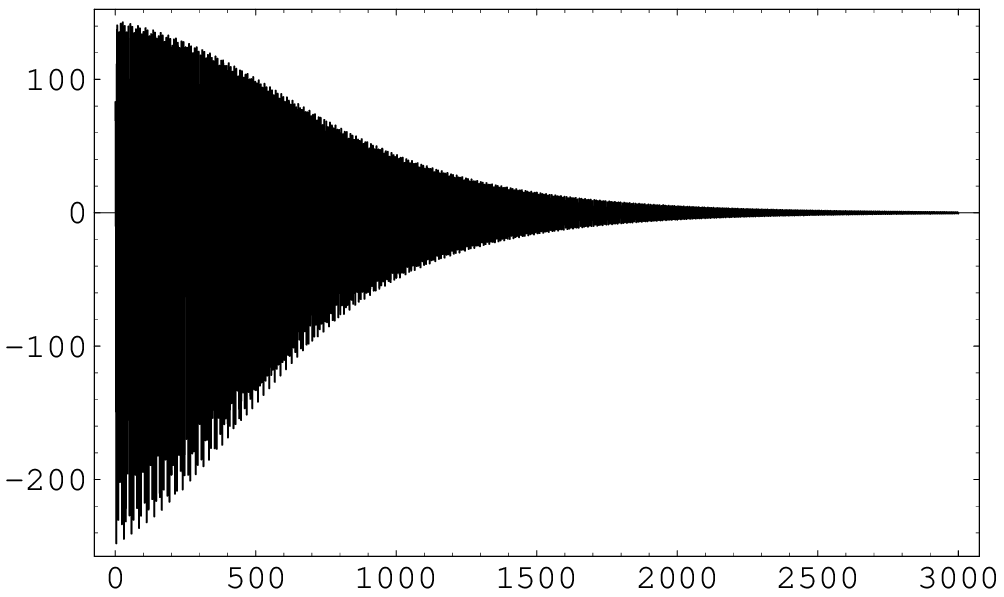}
\caption{Time series of $x$. $r=24$, $\sigma=10$, $b=8/3$.}
\label{timeser1}
\end{center}
\end{minipage}
\hfil
\begin{minipage}[b]{0.47\linewidth} 
\begin{center}
\epsfysize=100pt\epsfbox{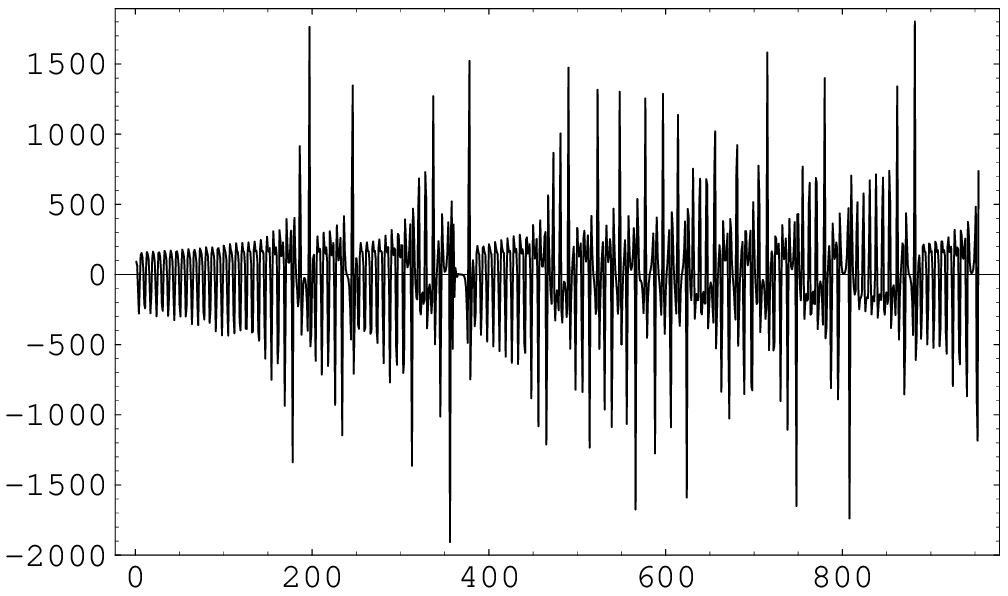}
\caption{Time series of $x$. $r=25$, $\sigma=10$, $b=8/3$.} 
\label{timeser2}
\end{center}
\end{minipage}
\hfil
\end{figure}

We consider a convection loop of water which has the Plandtl number $\sigma=10$  and the geometrical factor $b=8/3$. The threshold $r_t$ for these values is 24.7368\cite{Wig}. The time series of $x$ for $r_t=24$ is shown in Figure \ref{timeser1} and
that for $r_t=25$ is shown in Figure \ref{timeser2}. We find clearly that the system converges to a fixed point when $r<r_t$ but intermittency occurs when $r>r_t$.

\begin{figure}[b]
\begin{minipage}[b]{0.47\linewidth} 
\begin{center}
\epsfysize=100pt\epsfbox{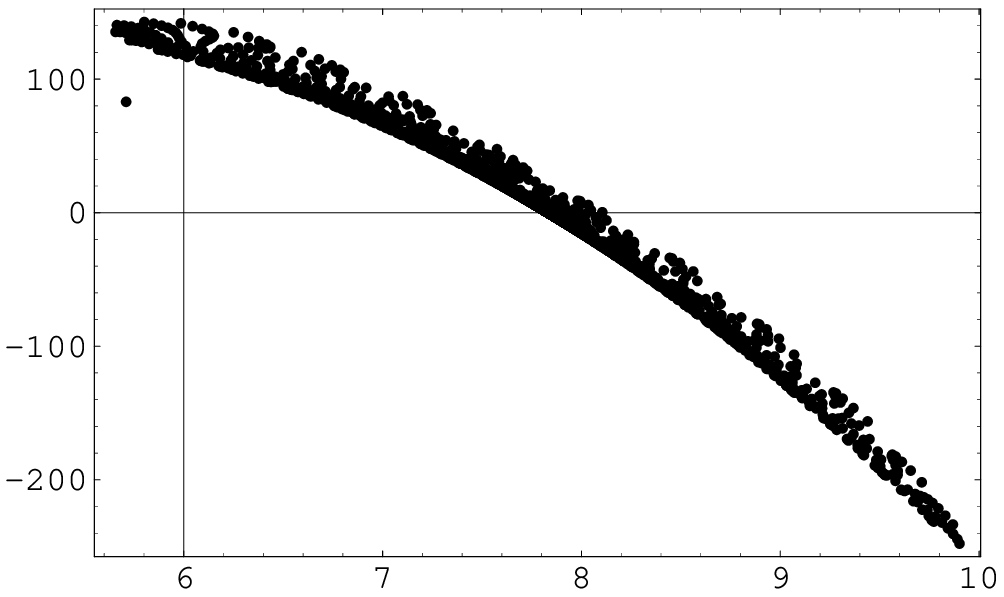}
\caption{Scatter plot in X-Z plane. $r=24$, $\sigma=10$, $b=8/3$.}
\label{timeser3}
\end{center}
\end{minipage}
\hfil
\begin{minipage}[b]{0.47\linewidth} 
\begin{center}
\epsfysize=100pt\epsfbox{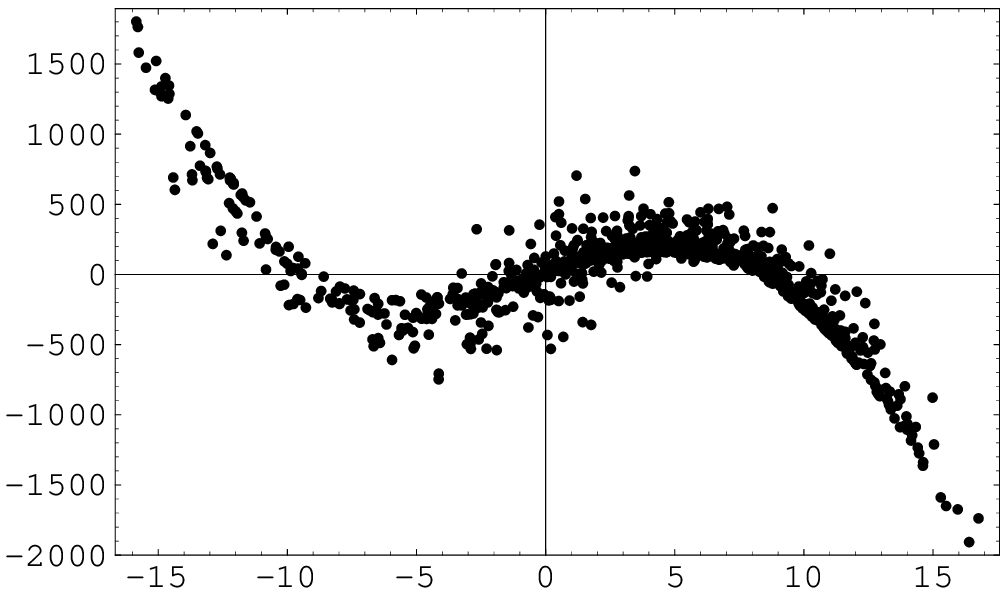}
\caption{Scatter plot in X-Z plane. $r=25$, $\sigma=10$,  $b=8/3$.} 
\label{timeser4}
\end{center}
\end{minipage}
\hfil
\end{figure}

After the contact transformation the scatter plot in the $X-Z$ plane is shown in Figures \ref{timeser3} and \ref{timeser4}.


We remark that in the X-Z plane, the scatter plot $X(Z)$ changes from the monotonic function in the case of $r<r_t$ to 3-valued function in the case of $r>r_t$.

\section{Convection loop model}
We consider the stability of thermosyphon or circulation of water in convection loop with heat source at the bottom and cooler at the top\cite{Cre}.
At the cooler, modulation of temperature in the form $\rho_0+\rho_1 \cos\omega t$ is assumed. The modified Lorenz equation is
\begin{eqnarray}
\frac{d x}{d t}&=&\sigma(y-x)\label{mlor1}\\
\frac{d y}{d t}&=&-y+(\rho_0+\rho_1\cos\omega t)x-xz\label{mlor2}\\
\frac{d z}{d t}&=&-bz+xy\label{mlor3}
\end{eqnarray}
Due to $\rho_1$ term, quasi-stable orbit can appear even when $\rho_0>r_t$.

The derivative $\displaystyle\frac{dZ}{dt}=f(X,Y,Z)$ after the contact transformation is
\begin{eqnarray}
f(X,Y,Z,t)&=&(YZ+\sigma Y^2+Y^2-\sigma X Z-X Z-X^3Y-\sigma X^4\nonumber\\
&&-bX Z-\sigma b X Y+\sigma b (\rho_0+\rho_1\cos\omega t) X^2-b X Y-\sigma b X^2)/X\nonumber\\
&&-\sigma(\rho_1\omega\sin\omega t) X, 
\end{eqnarray}

\subsection{Scatter plot in the $\Vec R$$^3$ and in the $X-Z$ plane}
\begin{figure}[b]
\begin{minipage}[b]{0.47\linewidth} 
\begin{center}
\epsfysize=100pt\epsfbox{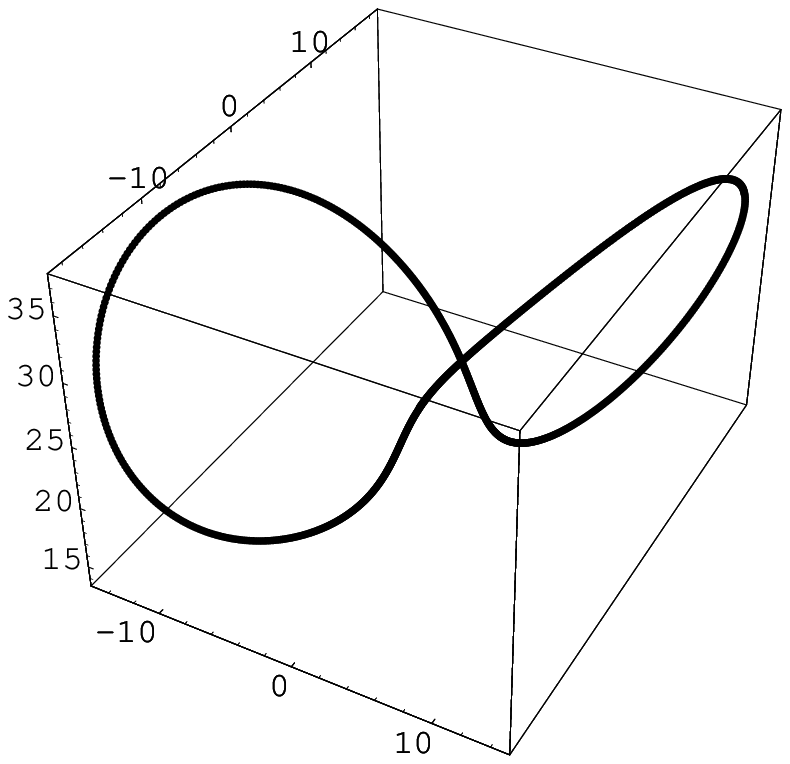}
\caption{Perturbed Lorenz model. $\rho_0=26$, $\rho_1=2.5$, $\omega=9$.}
\label{lorfig1}
\end{center}
\end{minipage}
\hfil
\begin{minipage}[b]{0.47\linewidth} 
\begin{center}
\epsfysize=100pt\epsfbox{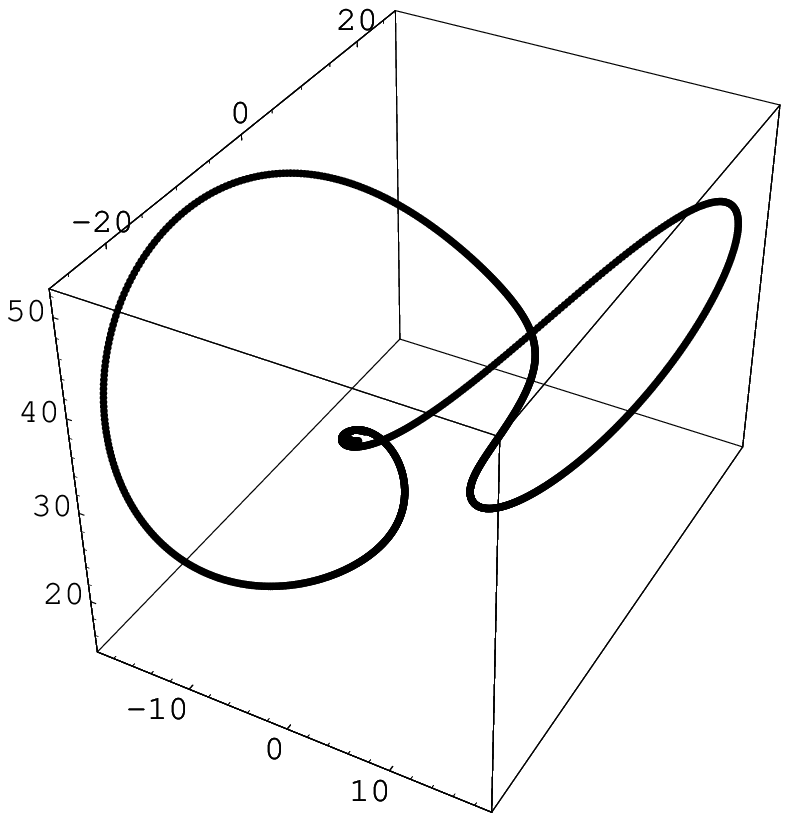}
\caption{Perturbed Lorenz model. $\rho_0=28$, $\rho_1=11$, $\omega=8.5$.} 
\label{lorfig2}
\end{center}
\end{minipage}
\hfil
\end{figure}

\begin{figure}[b]
\begin{minipage}[b]{0.47\linewidth} 
\begin{center}
\epsfysize=100pt\epsfbox{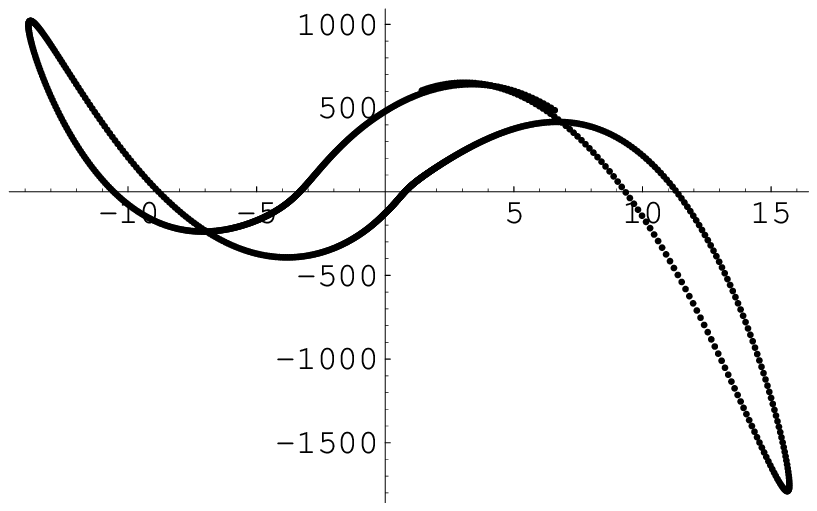}  
\caption{Scatter plot in $X-Z$ plane. $\rho_0=26$, $\rho_1=2.5$, $\omega=9$, $4750 \Delta t<t<5850 \Delta t$, $\Delta t=0.0015$} 
\label{lorfig26}
\end{center}
\end{minipage}
\hfil
\begin{minipage}[b]{0.47\linewidth} 
\begin{center}
\epsfysize=100pt\epsfbox{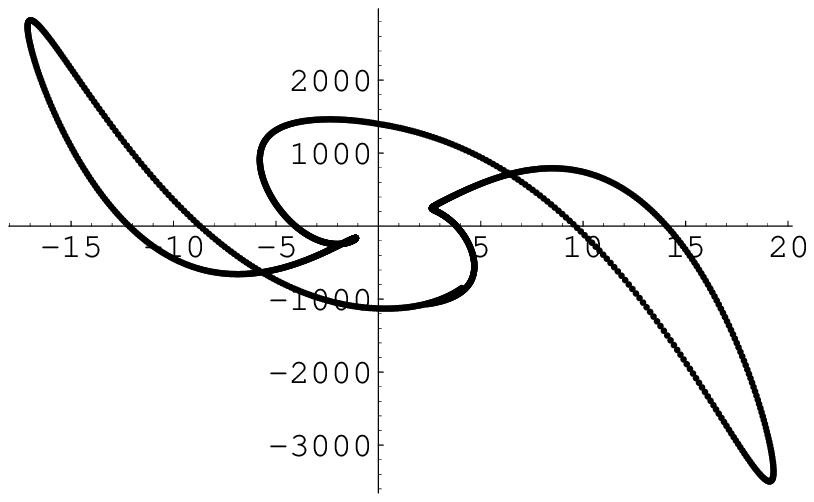}  
\caption{Scatter plot in $X-Z$ plane. $\rho_0=28$, $\rho_1=11$, $\omega=8.5$, $4000\Delta t<t<5600\Delta t$, $\Delta t=0.001$} 
\label{lorfig28}
\end{center}
\end{minipage}
\hfil
\end{figure}

In Figures \ref{lorfig1} and \ref{lorfig2}, we show quasi-periodic flow in ${\Vec R}^3$ for $\rho_0=26, \rho_1=2.5, \omega=9$ and for $\rho_0=28, \rho_1=11, \omega=8.5$, respectively. 
After the contact transformation, the corresponding quasi-periodic flows in the X-Z plane are shown in Figures \ref{lorfig26} and \ref{lorfig28}, respectively.   

\begin{figure}[b]
\begin{minipage}[b]{0.47\linewidth} 
\begin{center}
\epsfysize=100pt\epsfbox{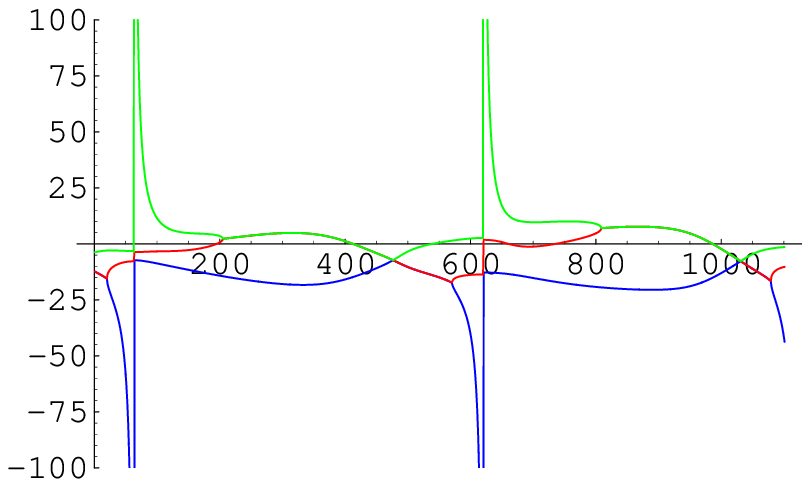}
\label{branch26}
\end{center}
\end{minipage}
\hfil

\begin{minipage}[b]{0.47\linewidth} 
\begin{center}
\epsfysize=100pt\epsfbox{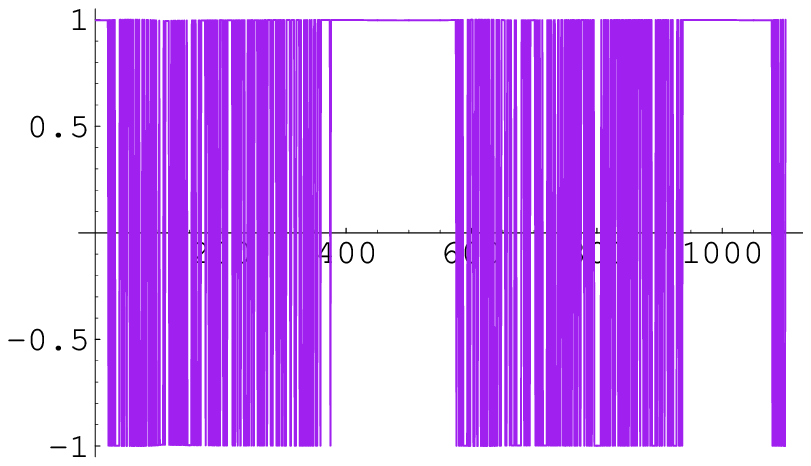}
\label{eigv26a}
\end{center}
\end{minipage}
\begin{minipage}[b]{0.47\linewidth} 
\begin{center}
\epsfysize=100pt\epsfbox{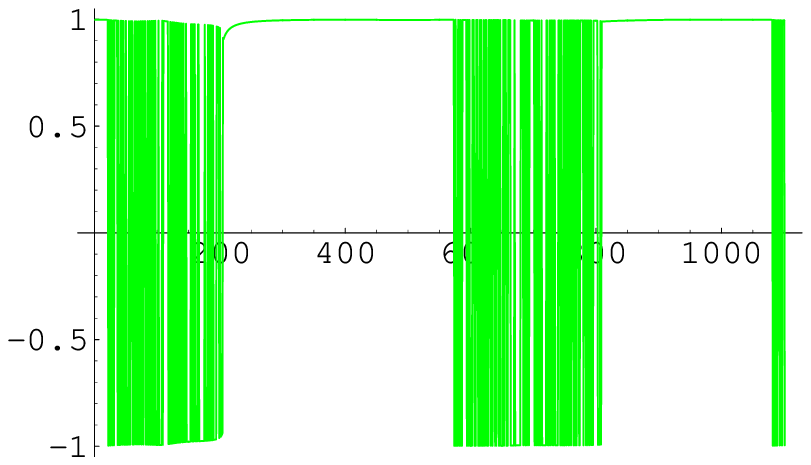}
\end{center}
\end{minipage}
\caption{Time series of the real part of the eigenvalues of Jacobian(a), the third component of the eigenvector corresponding to the lowest eigenvalue of Jacobian(b) and that corresponding to the second lowest eigenvalue. $\rho_0=26$, $\rho_1=2.5$, $b=8/3$, $\omega=9$.}
\label{eigv26b}
\hfil
\end{figure}

\begin{figure}[b]
\begin{minipage}[b]{0.47\linewidth} 
\begin{center}
\epsfysize=100pt\epsfbox{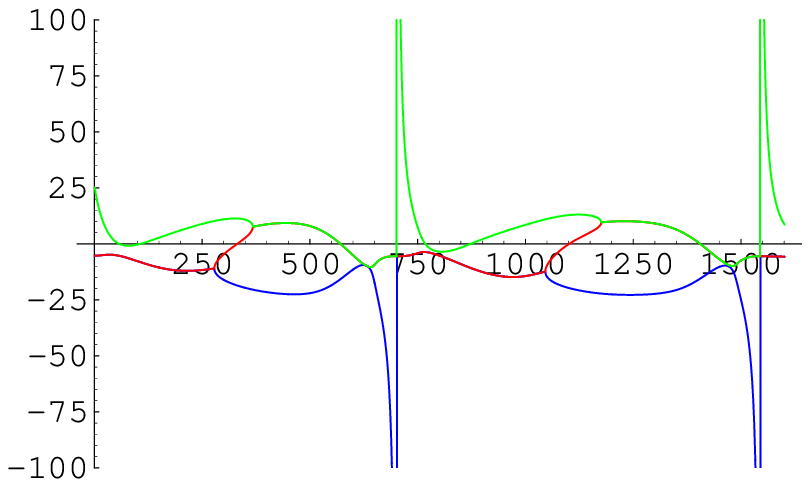}
\label{branch28}
\end{center}
\end{minipage}
\hfil

\begin{minipage}[b]{0.47\linewidth} 
\begin{center}
\epsfysize=100pt\epsfbox{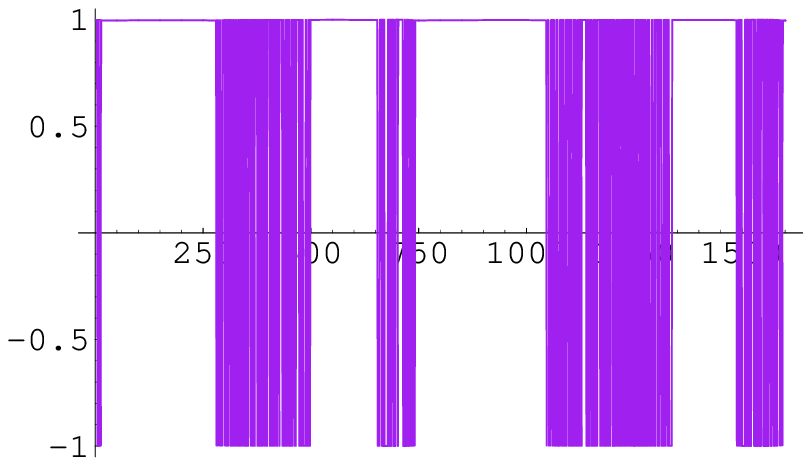}
\label{eigv28a}
\end{center}
\end{minipage}
\begin{minipage}[b]{0.47\linewidth} 
\begin{center}
\epsfysize=100pt\epsfbox{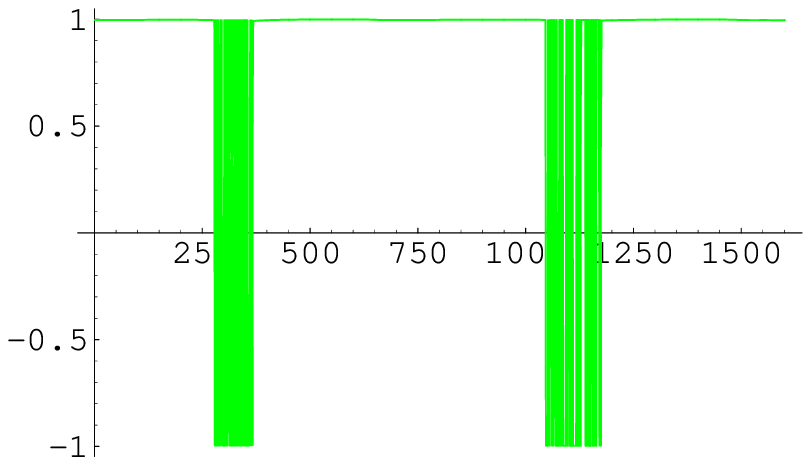}
\end{center}
\end{minipage}
\caption{Time series of the real part of the eigenvalues of Jacobian(a), third component of the eigenvector corresponding to the lowest eigenvalue of Jacobian(b), and that corresponding to the second lowest eigenvalue. $r=28$, $\rho_1=11$, $b=8/3$, $\omega=8.5$.}
\label{eigv28b}
\hfil
\end{figure}

We assign eigenvalues of the Jacobian (\ref{jacobi}) as $\lambda_i$, i=1,2,3 such that ${\rm Re } \lambda_1\leq
{\rm Re } \lambda_2\leq {\rm Re } \lambda_3$. 
The branching of the eigenvalues in the quasi-periodic region are shown in Figure \ref{eigv26b}(a), \ref{eigv28b}(a), for the small $\rho_1=2.5$ system and large $\rho_1=11$ system, respectively. 
We observe that, when the steps are started from near the origin, three real roots appear in the beginning but subsequently complex conjugate pairs corresponding to rotation around two fixed points dominate. In the plot of time-series of eigenvalues, there appears a discontinuity when $X=0$.  They are absent when contact transformation is not performed and do not affect the measurement of relative torsion number.

\subsection{Topological analysis}

For a dynamical system defined by
\begin{equation}
\frac{d{\Vec x}}{dt}={\Vec F}({\Vec x},t,\mu)
\end{equation}
where $\mu$ is the bifurcation parameter, one defines eigenvalues of
\begin{equation}
S(t,0)=exp[\int_0^t \frac{\partial{\Vec F}}{\partial {\Vec x}}({\Vec x}_0(t'),\mu)dt']
\end{equation}
as $\lambda_i, (i=1,3)$, such that@$|\lambda_1|\leq |\lambda_2|\leq |\lambda_3|$, and eigenvectors associated with them as ${\Vec e}_i, (i=1,3)$.
Around the periodic orbit 
\begin{equation}
{\mathcal C}_0=\{{\Vec x}_0(t)|0\leq t\leq T\},
\end{equation}
one defines two closed orbits\cite{UeAi}
\begin{equation}
{\mathcal C}_{i\zeta}=\{{\Vec x}_0(t)+\zeta {\Vec w}_i(t)|0\leq t\leq mT\}
\end{equation}
where $\zeta$ is a sufficiently small number and ${\Vec w}_i$ is the normalized eigenvector
\begin{equation}
{\Vec w}_i(t)={\Vec e}_i(t)/\|{\Vec e}_i(t)\|.
\end{equation}
The parameter $m$ is 1 when $\lambda_i>0$ and $2$ when $\lambda_i<0$.

In the system after the contact transformation, we define the normalized eigenvectors of Jacobian $\frac{\partial{\Vec F}}{\partial {\Vec X}}$, where ${\Vec X}=(X,Y,Z)$, but not that of $S(t,0)$ as ${\Vec w}_i$.

The local torsion number $r({\Vec w})$, which provides information on the rotation number of ${\mathcal C}_{i,\zeta=1}$ around ${\mathcal C}_0$ is defined by using
$\displaystyle {\Vec f}_1=\frac{\dot{\Vec X}}{\|\dot{\Vec X}\|}$,
$\displaystyle {\Vec f}_2=\frac{\dot{\Vec X}\times \ddot{\Vec X}}{\|\dot{\Vec X}\times \ddot{\Vec X} \|}$,
$\displaystyle {\Vec f}_3={\Vec f}_1\times{\Vec f}_2$ and
$\alpha_j(t)={\Vec w}(t)\cdot{\Vec f}_j$
as
\begin{equation}
r({\Vec w})=\frac{1}{2\pi}\int_0^T \frac{\alpha_2(t)\cdot\frac{d}{dt}\alpha_3(t)-
\frac{d}{dt}\alpha_2(t)\cdot\alpha_3(t)}{\alpha_2(t)^2+\alpha_3(t)^2}
\end{equation}

In Figure \ref{eigv26b}(b,c), and  \ref{eigv28b}(b,c) we show the third component of the two normalized eigenvectors ${\Vec w}_1$ and ${\Vec w}_2$ for the small $\rho_1$ system and  large $\rho_1$ system, respectively.
The horizontal lines of the third component of the eigenvectors ${\Vec w}_1$ and ${\Vec w}_2$ equal 1 indicate stability of the orbit $C_0$. In the case of small $\rho_1$ (Figure \ref{eigv26b}(b)), we find two stable windows, which indicates LR type period 2 structure. While in the case of large $\rho_1$ (Figure \ref{eigv28b}(b)), there is an additional breaking in the eigenvector ${\Vec w}_1$  which corresponds to a compensating twist on the LR type period 2 structure.

The eigenvector ${\Vec w}_3$ has also small first and second components, and its third component oscillates between -1 and +1. We measure the relative torsion number of the orbit $C_{3,1}$, since ${\Vec w}_1$ and ${\Vec w}_2$ specify the trajectory of $C_0$. We observed that the local torsion number $r({\Vec w}_3)$ in the case of small $\rho_1$ is -1/2 and that of large $\rho_1$ is 1/2. 

In order to get local crossing number $C_i$ of the orbit\cite{ArMo}, we measure the position of the highest peak in the power spectrum of the orbit of Figure\ref{lorfig26} and Figure\ref{lorfig28}. 
The results of small $\rho_1$ and large $\rho_1$ together with the pure chaotic case without perturbation ($\rho_1=0$) are shown in Figures 11,12 and 13,14
 respectively.
\quad Templates are characterized in terms of number of half twists $\xi$ and the direction of folding $p\in \{-1,1\}$, as $(\xi,p)$, and the local crossing number $C_{\langle 2\rangle}=2\xi+p$ in the case of period doubling bifurcation\cite{ArMo}.

In the power spectrum of $\rho_0=26$ and $\rho_1=2.5$, $\omega_1=9$, we observe the peak position is at $\omega=4$ in arbitrary unit. When the perturbation is set to $\rho_1=0$, the position moves to $\omega=3$. Hence, using the information $r({\Vec w}_3)=-1/2$, i.e. $p=-1$, we assign $\displaystyle \frac{C_2}{2^2}=\frac{3}{4}$. Since $C_2=3$, the template of small $\rho_1$ is $(\xi,p)=(2,-1)$.
 
In the power spectrum of $\rho_0=28$ and $\rho_1=11$, $\omega_2=8.5$, the peak position at $\omega=4$ and it does not change after setting $\rho_1=0$.  Using the information $r({\Vec w}_3)=1/2$, i.e. $p=1$, we assign $\displaystyle {C_1}=\frac{1}{1}$. The template of large $\rho_1$ is $(\xi,p)=(0,1)$.

\begin{figure}[htb]
\begin{minipage}[b]{0.47\linewidth} 
\begin{center}
\epsfysize=125pt\epsfbox{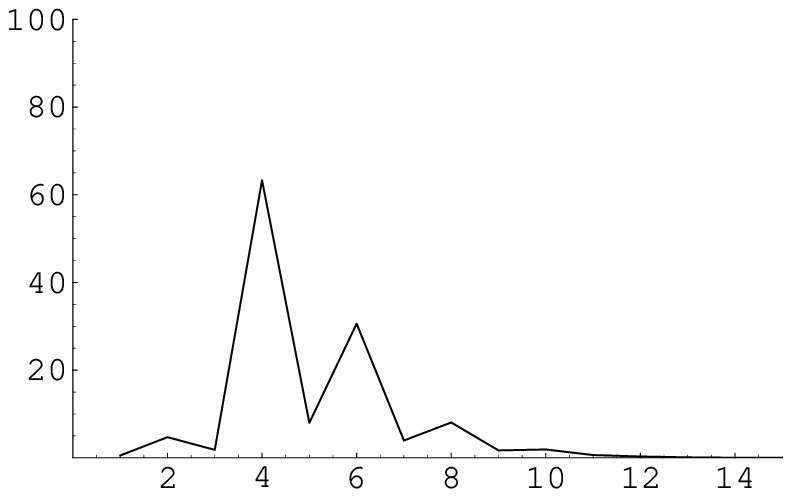}\label{power1a}
\caption{The power spectrum of $\rho_0=26,\rho_1=2.5,\omega=9$.}
\end{center}
\end{minipage}
\hfil
\begin{minipage}[b]{0.47\linewidth} 
\begin{center}
\epsfysize=125pt\epsfbox{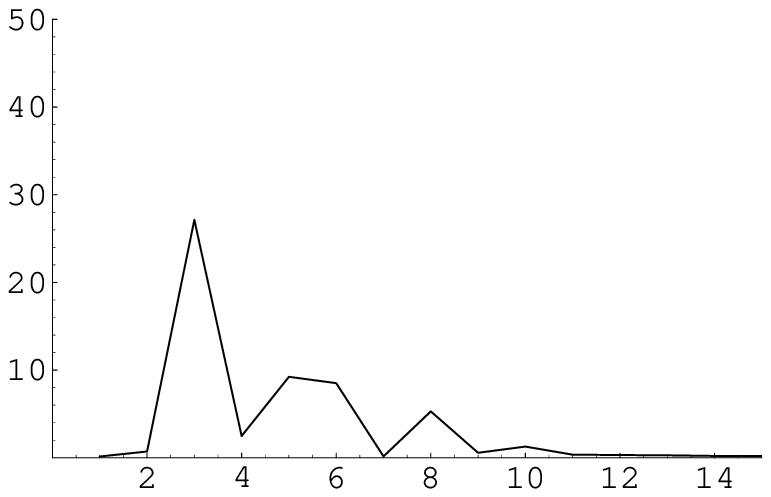}\label{power1b}
\caption{The power spectrum of $\rho_0=26,\rho_1=0$.}
\end{center}
\end{minipage}
\hfil
\end{figure}

\begin{figure}[htb]
\begin{minipage}[b]{0.47\linewidth} 
\begin{center}
\epsfysize=125pt\epsfbox{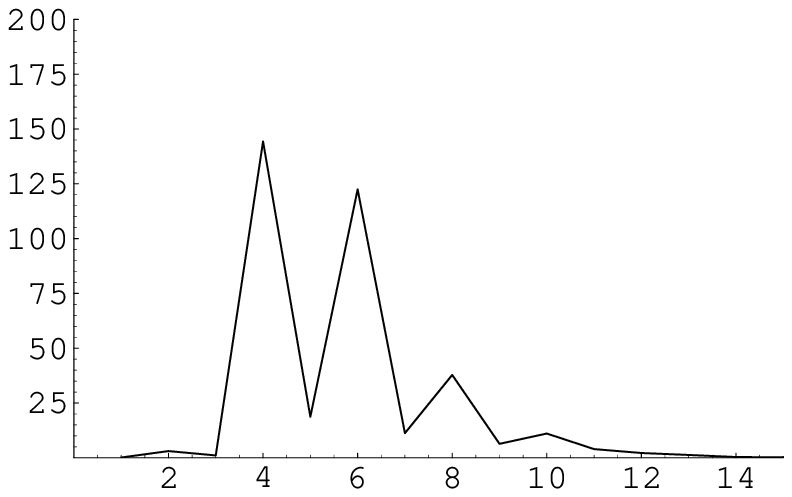}\label{power2a}
\caption{The power spectrum of $\rho_0=28,\rho_1=11,\omega=8.5$.}
\end{center}
\end{minipage}
\hfil
\begin{minipage}[b]{0.47\linewidth} 
\begin{center}
\epsfysize=125pt\epsfbox{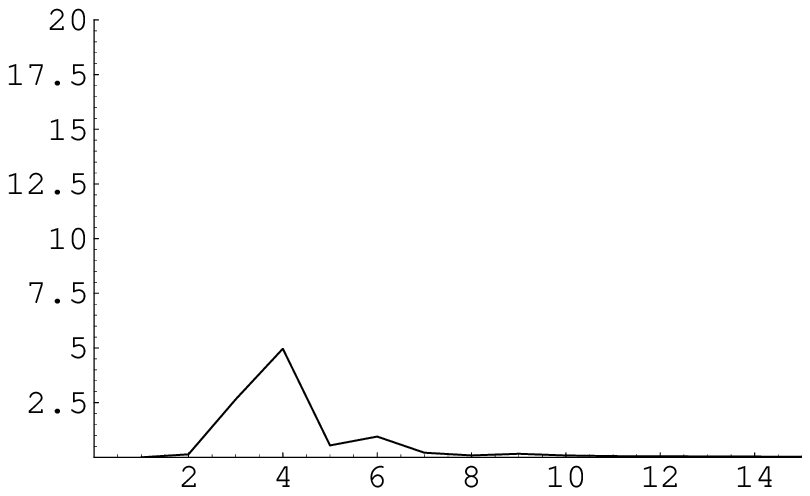}\label{power2b}
\caption{The power spectrum of $\rho_0=28,\rho_1=0$.}
\end{center}
\end{minipage}
\hfil
\end{figure}

\section{Single mode laser}
In 1975, Haken showed that under certain conditions a coherently pumped homogeneously broadened ring laser obeys the Lorenz equation\cite{Hak,Hakb}.  Whether laser actually behaves in this way was studied, and data of a kind of infrared single-mode gas laser was simulated\cite{Lug}.  

In this model, the inversion density is defined as $D$, the electric field and polarization are specified as $\tilde x=x e^{i\theta \tilde\kappa t}$ and $\tilde p=p e^{i\Delta t}$, respectively. In the single mode laser, $\Delta=\theta\tilde\kappa$, and the coupled equation of the system with external static electric field $y$ becomes
\begin{equation}
\frac{dx}{dt}=-\tilde\kappa(i\theta x+x-y+2Cp)
\end{equation}
\begin{equation}
\frac{d p}{d t}=xD-(1+i\Delta)p 
\end{equation}
\begin{equation}
\frac{d D}{d t}=-\tilde\gamma[\frac{1}{2}(xp^*+x^*p)+D+1] 
\end{equation}

We consider the modulation $g=e^{i\Delta t}$ as a structure group $U(1)$ defined on the base space.  The complex space specified by $x, p$ and $D$ is projected onto ${\Vec R}^3$ specified by $|x|, |p|$ and $D$ by identifying all the fields $x$ and $p$ that are related as $g^{-1}x g=x'$. 

The stable solution that is obtained by taking l.h.s. of the equations to be zero gives the relation\cite{Lug} 
\begin{equation}
y=|x|[(1-\frac{2C}{1+\Delta^2+|x|^2})^2+(\theta+\frac{2C\Delta}{1+\Delta^2+|x|^2})^2]^{1/2}
\end{equation}

\begin{figure}[b]
\begin{center}
\begin{minipage}[b]{0.47\linewidth} 
\epsfysize=100pt\epsfbox{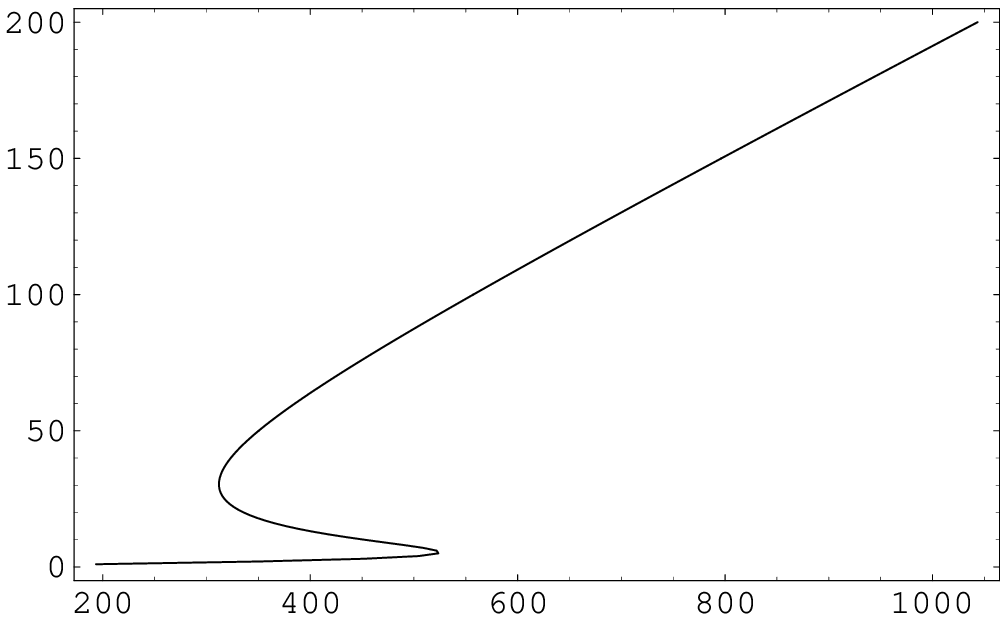}
\caption{The electric field $|x|$ of the stable or quasi stable orbit as a function of the external static field $y$. }
\label{laserxy}
\end{minipage}
\end{center}
\end{figure}

The function $|x|(y)$ shows a bending behaviour and in the region $300 < |x| < 500$ it is 3-valued function(Fig.\ref{laserxy}).

On the projected space we can perform the contact transformation as before and we obtain
\begin{equation}
\frac{dX}{dt}=Y 
\end{equation}
\begin{equation}
\frac{d Y}{d t}=Z 
\end{equation}
\begin{eqnarray}
\frac{d Z}{d t}&=&-i\tilde\kappa \theta Z -\tilde\kappa Z\nonumber\\
&&-2\tilde\kappa C[Y F_1+X(-\tilde\gamma(\frac{1}{2}(X F_2^*+X^* F_2)+F_1+1))]\nonumber\\
&&-(Z+i\tilde\kappa \theta Y+\tilde\kappa Y)(1+i\Delta) \label{dzdt}
\end{eqnarray}
where
\begin{equation}
F_1=\frac{-Z-i\tilde\kappa\theta Y-\tilde\kappa Y -(Y+i\tilde\kappa\theta X+\tilde\kappa X-\tilde\kappa y)(1+i\Delta)}{2\tilde\kappa C X}
\end{equation}
\begin{equation}
F_2=\frac{-Y-i\tilde\kappa\theta X-\tilde\kappa X+\tilde\kappa y}{2\tilde\kappa C}
\end{equation}

\subsection{Scatter plot in the $|X|-|Z|$ plane}
Due to the positive real $y$, the system is unstabilized and shows a strong sensitivity to the initial condition. In this paper, we present the result of $\tilde\kappa=1$, $C=500$, $\Delta =\theta=5$, and the initial condition $X=350, Y=0, Z=1$.
The scatter plot of $|Z|$ in the  $|X|-|Z|$ plane for each $y$ from 230 to 340 are plotted in Figure 17 
for $y\geq 270$ and Figure 16 
for $y\leq 260$.
Convergence to a stable branch can be observed for $y\geq 300$, but at $y=270$ intermittency occurs and the situation becomes chaotic for $y\leq 260$.

Although the chaotic behaviour in the time series of $x$ can be observed by the bursting and spiking phenomena\cite{Lug}, the scatter plot in the $|X|-|Z|$ plane shows clearer transition between chaos and quasi-stable mode. We observe intermittency begins to occur at around $y=270$ and the system become chaotic for smaller $y$ values.


\begin{itemize}
\item Fig.16 Scatter plot of $(|X|,|Z|)$ for $y\leq 260$. The points correspond to $y$=230, 240, 250, 260, from left to right. See the extra file:laser1.gif
\item Fig.17 Scatter plot of $(|X|,|Z|)$ for $y\geq 270$. The points correspond to $y$=270, 280, $\cdots$, 340, from left to right. See the extra file:laser.gif
\end{itemize}

\setcounter{figure}{17}
\subsection{Intermittency}
\begin{figure}[b]
\begin{minipage}[b]{0.47\linewidth} 
\begin{center}
\epsfysize=120pt\epsfbox{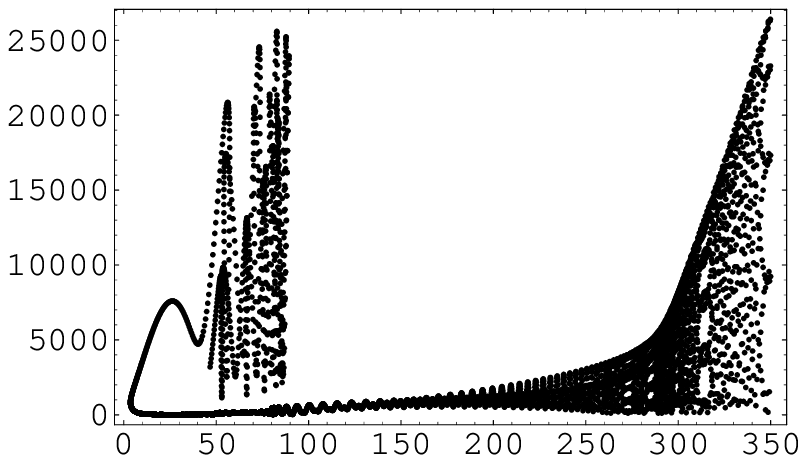}
\caption{Scatter plot of $(|X|,|Z|)$ in the case of $y=270$.
}
\label{int270}
\end{center}
\end{minipage}
\hfil
\begin{minipage}[b]{0.47\linewidth} 
\begin{center}
\epsfysize=120pt\epsfbox{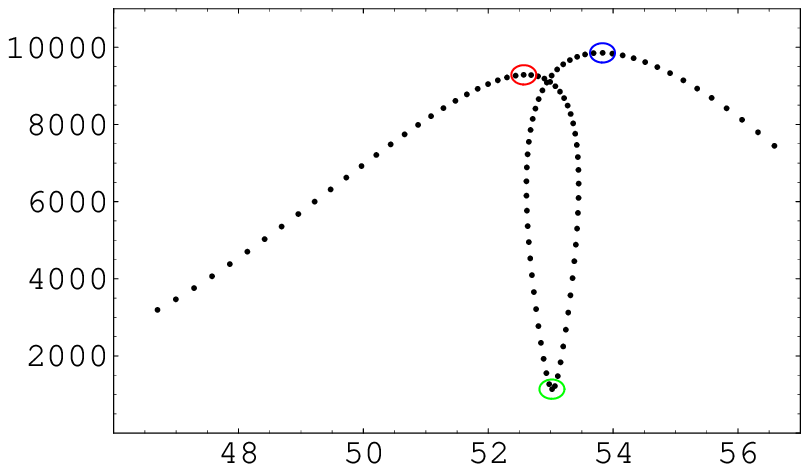}
\caption{Scatter plot of $(|X|,|Z|)$ in the case of $y=270$  near $|X|=53$.
The right local maximum, local minimum and the left local maximum are assigned by circles.}
\label{int270part}
\end{center}
\end{minipage}
\end{figure}

The scatter plot of $(|X|,|Z|)$ at $y=270$ is presented in Figure \ref{int270}.   The intermittency near $|X|=53$ is shown in Figure \ref{int270part}.

\begin{figure}[b]
\begin{minipage}[b]{0.47\linewidth} 
\begin{center}
\epsfysize=200pt\epsfbox{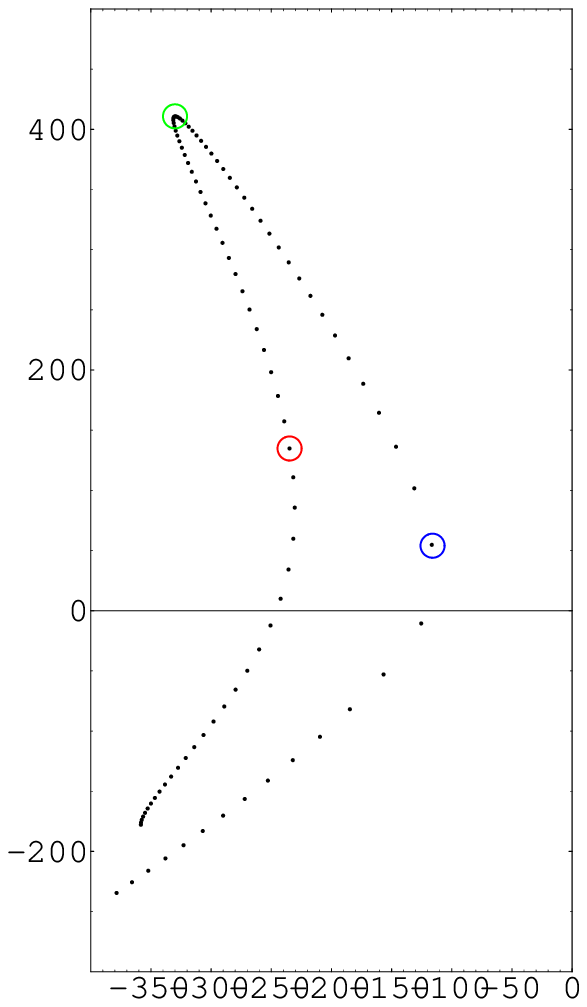}
\caption{Scatter plot of complex eigenvalues of the Jacobian. Eigenvalues of boomerang type which have negative real part. Ordinate is the real part and abscissa is the imaginary part. The three circles from right to left of Figure \ref{int270part} correspond to the circle in the right, in the top and in the left, respectively.}
\label{lam1}
\end{center}
\end{minipage}
\hfil
\begin{minipage}[b]{0.47\linewidth} 
\begin{center}
\epsfysize=200pt\epsfbox{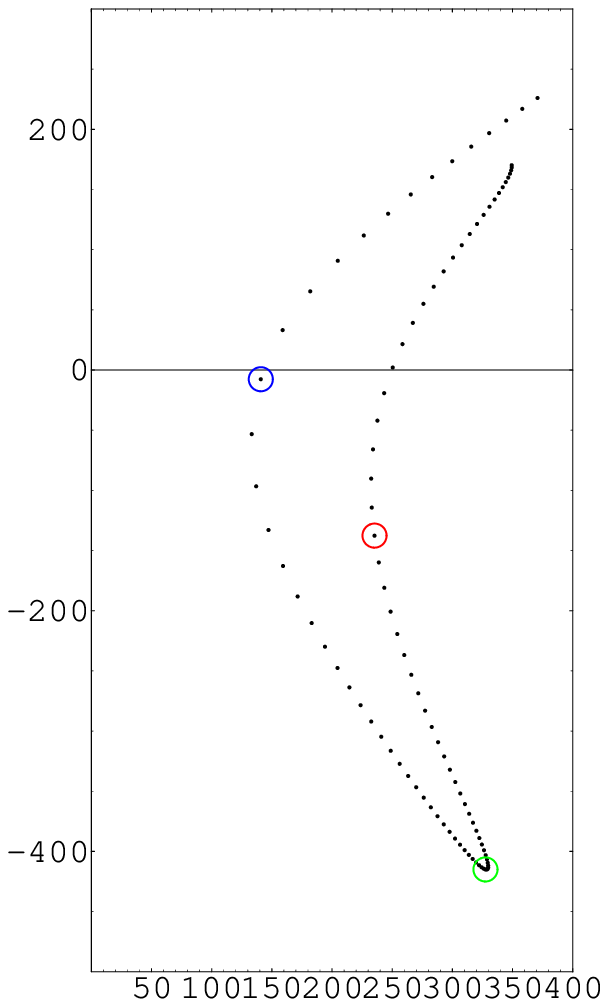}
\caption{Scatter plot of complex eigenvalues of the Jacobian. Eigenvalues of boomerang type which have positive real part. The three circles from right to left of Figure \ref{int270part} correspond to the circle in the left, in the bottom and in the right, respectively.}
\label{lam3}
\end{center}
\end{minipage}
\end{figure}

\begin{figure}[b]
\begin{minipage}[b]{0.47\linewidth} 
\begin{center}
\epsfysize=120pt\epsfbox{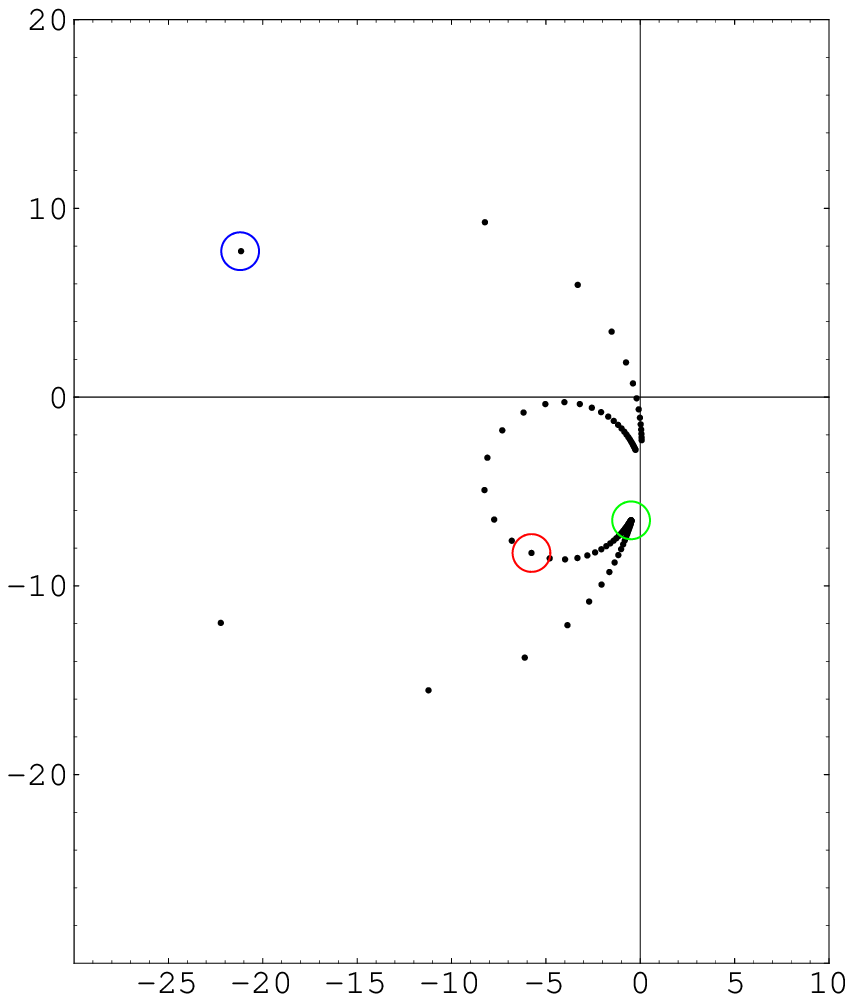}
\caption{Scatter plot of ring type complex eigenvalues of the Jacobian.  
The three circles from right to left of Figure \ref{int270part} correspond to the circle in the 2nd quadrant, at the cusp and in the 3rd quadrant, respectively.}
\label{lam2}
\end{center}
\end{minipage}
\end{figure}

In order to check the type of intermittency, we measure the time series of the eigenvalues of the Jacobian. 
 Figures \ref{lam1}, \ref{lam3} and \ref{lam2} are the scatter plot of the time series of the three complex eigenvalues. There appears a pair of boomerang type and a ring type. 

As the point $(|X|,|Z|)$ moves from right, makes a loop, and then move to left as shown in Figure \ref{int270part}, the eigenvalues
of ring type move from around $(0,-3i)$ counter clock wise, move to around $(-0.5,-6.5i)$, make a cusp, and then move clock wise as shown in Figure \ref{lam2}. The movements of eigenvalues of boomerang type are counter clockwise.

As the eigenvalue of boomerang type approach the real axis,  $|Z|$ takes a local maxima, and as the eigenvalue of ring type makes a cusp, $|Z|$ takes a local minimum. These properties can be qualitatively understood as the movement along the trajectory of the $U(1)$ group makes a jump of the flow in $(|X|,|Y|,|Z|)$ base space, which has the multivalued structure as in the original Lorenz model.

\section{Discussion and outlook}
We applied the contact transformation to the Lorenz equation and its extension, and analyzed transition from chaos to quasi-periodic phase in convection loop and in the single mode laser.  The branching of eigenvalues of the Jacobian after contact transformation allows visualizing transition from orbit near one fixed point to the other one.

  In the case of convection loop we considered perturbation by $\rho_1 \cos\omega t$ term with large $\rho_1$ and small $\rho_1$.  From the behaviour of eigenvectors of the Jacobian, we observe transition from quasi-stable to unstable orbits. The relative torsion number can be measured by using the eigenvectors and the local crossing number can be measured from the power spectra of $Z(t)$ in periodic orbit with $\rho_1\cos \omega t$ type perturbation and without perturbation. The topological data after contact transformation are simpler than those before contact transformation. 
Without contact transformation, the measurement of relative torsion number $r_i$ is relatively ambiguous since $r_i$ for $x,y$ and $z$ directions are equivalent and there is no guideline for selecting one like $Z$.  

The $X-Z$ plot in the convection loop and $|X|-|Z|$ plot in the single mode laser allows to visualize characteristic pattern change when the system changes from chaotic to quasi-periodic phase.

We observed intermittency in the single mode laser system at around $y=270$ and observed correlation between the critical movement of complex eigenvalues of the Jacobian and the sudden large $|Z|$ oscillation. The orbit in the fiber space specified by the $U(1)$ group causes a jump in  the trajectory in the base space
specified by $(|X|,|Y|,|Z|)$. It is a new type of intermittency.

The topological characteristics of the orbit can be attributed to the jump in the relative linking number of the orbit ${\Vec x}_0+{\Vec w}_3$ around the quasi-periodic orbit ${\Vec x}_0$, where ${\Vec w}_3$ is the normalized eigenvector of the Jacobian, associated with the eigenvalue possessing the largest real part. We observe that the behaviour of the eigenvectors corresponding to eigenvalues with the smallest and the second smallest real part give information of the template and the orbit $C_0$. 

When the Lyapnov exponents satisfy $\lambda_1>\lambda_2=0>\lambda_3>\cdots >\lambda_d$ and there is a strong local attraction, the Lyapnov dimension becomes
\begin{equation}
d_L=2+\frac{\lambda_1}{|\lambda_3|}<3
\end{equation}
which means that the system can be mapped on 2-dimensional template\cite{KY,BP}. In the decomposition of the three dimensional space to two dimension + one dimension space, the 2-dimensional template could always contain stable manifold. 

In higher dimensional space, 4-dimension as an example, the chosen 2-dimensional template could contain only the unstable manifold. To study it, we consider
 the extended Lorenz equation\cite{UeKu} in ${\Vec R}^4$.
\begin{eqnarray}
\frac{d x}{d t}&=&\sigma(y-x)\nonumber\\
\frac{d y}{d t}&=&-y+rx-x(z+w)\nonumber\\
\frac{d z}{d t}&=&-bz+xy\nonumber\\
\frac{d w}{d t}&=&kx^2-cw
\end{eqnarray}
where parameters are $\sigma=16, b=4, c=1, k=0.02$.

In this model, fixed points exist at $\displaystyle x=y=\pm\sqrt{\frac{r-1}{k+1/b}}$ and a pitchfork bifurcation takes place for the parameter $r\simeq 530$\cite{UeKu}. 
Around this point $kx^2-cw$ becomes positive and $w$ increases monotonically but in the projected $(x,y,z)$ space we can observe the attractor.
 
When we perform the contact transformation
\begin{equation}
\frac{d X}{d t}=Y, \quad\frac{d Y}{d t}=Z, \quad
\frac{d Z}{d t}=W \nonumber
\end{equation}
\begin{eqnarray}
\frac{d W}{d t}&=&(Z(Y+Z)^2-\sigma(W X^2((1+b+c)X-2Y)+cX^5 Y-c X^2Y^2+X^4Y^2\nonumber\\
&+&2XY^3+cX^3Z+X^5Z-2X^2YZ-c X^2YZ-2Y^2Z\nonumber\\
&+&2XY^2Z-2YZ^2+bX^2(-Y^2+XZ-YZ+cX(Y+Z))\nonumber\\
&-&\sigma^2(WX^3+bkX^6+2X^5Y+2kX^5Y-bX^2Y^2+2XY^3+bX^3Z-2X^2YZ\nonumber\\
&-&Y^2Z+cX^2((-1+r)X^2+X^4-Y^2+X(bY+Z)))/(\sigma X^3)
\end{eqnarray}
we find spiral divergence in the $Z-W$ plane.

The dynamical flows in these frames are trapped in their corresponding unstable manifolds. To remedy it, one can choose $w$ as $X$ and linearize $k x^2-c w$ as $k(x_0^2+x_0 \bar x)-cw$ where $x_0$ is the coordinate of the fixed point and study dynamics on the template in the co-dimensional space of $w$, i.e. $\bar x,y,z$ space. 
In general, when the dimension of the space is larger than three, it is necessary to choose coordinate in the contact transformation so that the template contains the 2-dimensional stable manifold.


I acknowledge with thanks that graduate students M. Horinouchi and N. Kamimura obtained some simulation data in this paper. Thanks are also due to the referee for bringing references\cite{AiUe,UeAi,Ue,UeKu} to my attention and Dr. T. Arimitsu for kindly sending me his references and helpful discussion.

\end{document}